\documentclass[aps,prb,preprint,showpacs,showkeys,floatfix]{revtex4}

\usepackage{graphicx,color}
\usepackage{multirow,slashbox}
\usepackage{natbib}

\graphicspath{{figs/}}
\begin{document}

\title{Buckling behavior of tenary one-dimensional van der Waals heterostructures}

\author{Jing Wan}
    \altaffiliation{Corresponding author: luckwan16@163.com}
    \affiliation{School of Mechanics and Safety Engineering, Zhengzhou University, Zhengzhou, People's Republic of China}

\author{Shu Lin}
    \affiliation{School of Mechanics and Safety Engineering, Zhengzhou University, Zhengzhou, People's Republic of China}





\date{\today}
\begin{abstract}
1D van der Waals heterostructures (1D vdWH) were recently reported to be successfully synthesized. We perform molecular dynamics simulations to investigate the buckling behavior of a 1D vdWH composed of inner carbon nanotube, a middle boron nitride nanotube and an outer molybdenum disulfide nanotube. We find that as the temperature increases, the 1D vdWH gradully loses its stability and the peak value of compressive stress decreases. Slenderness ratio have a slight influence on the strength and stability of 1D vdWH under axial compression.

\end{abstract}
\keywords{One-dimensional Heterostructure, Buckling Behavior, Molecular Dynamics}
\maketitle
\pagebreak

\section{Introduction}

Van der Waals heterostructures, created from various nanomaterials of different dimensions by van der Waals interaction, have attracted tremendous attention in the past years\cite{Novoselov2016}. Since the goodness of different ingredients is combined in one ultimate material, heterostructures have been proven to have unique electronic, mechancial and optical properties. These extraordinary properties make heterostructures promsing candidates for field-effect, light-emitting and photovoltaic devices. One-dimensional van der Waals heterostructures (1D vdWHs), a class of nanomaterials where different nanotubes are coaxially stacked, were recently reported to be successfully synthesized\cite{Xiang2020}. Subsequently, these coaxial heteronanotubes demonstrated their advantages in the practical applications\cite{Xiang2021}. Liu et al. found that the coaxial 1D vdWH comprised of a carbon nanotube (CNT) core and a thickness tuneable covalent organic framework shell could sever as an efficient metal-free oxygen electrocatalysts\cite{Liu2021}. Wang et al. experimentally measured the in-plane thermal conductivity of 1D vdWH consisting of inner single-walled CNT and outer boron nitride nanotube (BNNT), and found that the heterostructure has a great improvement in thermal conductivity than the bare single-walled carbon nanotube (SWCNT). 

Mechanical property has an important influence on the performance and service-life of the low-dimensional devices. The coaxial heterostructure composed of two nanotubes with different diameters is the most familiar 1D vdWHs, such as unary double-walled carbon nanotubes (DWCNTs). Past works have studied the mechancial properties of DWCNTs. Compared with CNTs, BNNTs have higher thermal stability in the air. Therefore, the outer BNNT usually is synthesized as a protective coating for inner CNT, ulimately forming the binary 1D vdWH consisting of a coaxial CNT inside a BNNT (CNT@BNNT). The molecular dynamics (MD) simulation results show that the outer BNNT has a better protective effect on the inner tube than does the outer CNT under high-temperature conditions\cite{Liew2011}. He et al. report that the CNT@BNNT possesses excellent mechanical strength and the calculated Young's modulus is as high as 0.856~TPa\cite{He2019}. Experimental measurement results indicate that the encapsulating with protective BNNTs would enable a large increase in the compressive strength of CNT arrays\cite{Jing2016}. However, the mechancial properties of the newly synthesized tenary 1vdWHs are still unclear. 

In the recent experimental discovery\cite{Xiang2020}, the typical tenary structure features small diameters of 4-5~nm and contains a inner CNT, a middle BNNT and an outer molybdenum disulfide nanotube (MoS$_2$NTM), and we focus on the mechancial properties of this CNT@BNNT@MoS$_2$NT 1vdW.

In the present study, we perform MD simulations to study the buckling behavior of tenary 1D vdWH composed of CNT, BNNT and MoS$_2$NT under axial compression. We find that with the temperature increase, the 1D vdWH loses its stability and the peak value of compressive stress decreases. Slenderness ratio have a slight influence on the strength and stability of 1D vdWH under axial compression.


\section{Models and methods}

\begin{figure}[tb]
  \begin{center}
    \scalebox{1}[1]{\includegraphics[width=16cm]{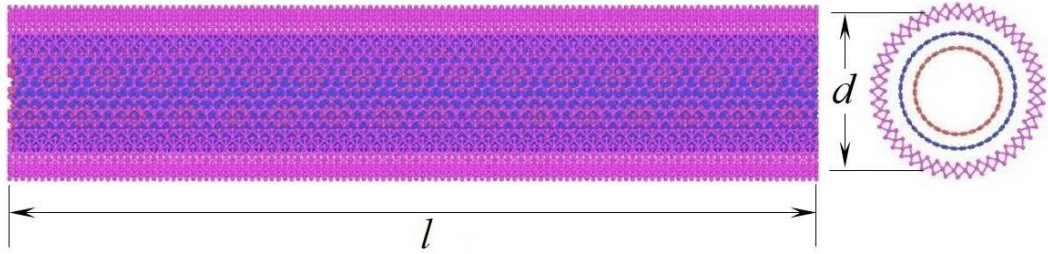}}
  \end{center}
  \caption{Schematic of 1D CNT@BNNT@MoS$_2$NT heterostructure under uniaxial loading. The CNT@BNNT@MoS$_2$NT heterostructure formed by three different shells: an inner CNT, a middle BNNT and an outer MoS$_2$NT.}
  \label{fig_structure}
\end{figure}

The CNT@BNNT@MoS$_2$NT heterostructure contains three different shells: an inner CNT, a middle BNNT and an outer MoS$_2$NT, as shown in Fig.~\ref{fig_structure}~(a). Each shell of this 1D vdWH is a single crystal and the interaction between shells is van der Waals force. The diameter $d$ of this 1D vdWH is about 4~nm and the length $l$ is about 25~nm. Small diameter of nanotube will bring large strain energy and cause structural instability, especially for the MoS$_2$NT who has a thickness of three atomic planes. Theoretical results show that when the diameter of MoS$_2$NT is larger than 3.7~nm, the strain energy reaches the minimum and the MoS$_2$NT gets structure stability\cite{Xiang2020}. For the 1D vdWH of the diameter $d$ around 5~nm adopted in our analysis, the corresponding diameter of its component are 2.71~nm for CNT(20, 20), 3.39~nm for BNNT(25, 25) and 4.78~nm for MoS$_2$NT(26, 26), respectively. 

The loading test is illustrated in Fig.~\ref{fig_structure}~(b). Under uniaxial compression, the bottom end of the 1D vdWH is fixed, and the upper end of the 1D vdWH moves gradully downward. The entire loading test stops when the structure occurs serious buckling. The stability test is simulated using MD approach. Specifically, the whole structure is reshaped with the steepest descending algorithm. After energy minimazition, the structure is relaxed within NVT ensemble, and then the atoms at both ends of the 1D vdWH are fixed. In the loading process, the upper end moves 0.01~nm downward and followed with 40~ps of relaxtion for the rest part of the structure each time untill the structure is crushed. 

All the MD simulations are performed by larger-scale atomic molecular massively parallel simulator (LAMMPS) package\cite{Plimpton1995}, while the OVITO tool is used for visualization\cite{Stukowski2009}. The interatomic interaction among Mo and S atoms in MoS$_2$NT is evaluated by Stillinger-Weber potential parameterized by Jiang\cite{Jiang2015}, which has been successfully used in predicting properties of MoS$_2$ nanosheet. Tersoff potential is used to descirbe the interatomic interaction among B and N atoms in BNNT, while the interaction between C atoms in the CNT is controlled by AIREBO potential. The van der Waals interlayer interaction is described by using a 12-6 Lennard-Jones (LJ) potential. The parameters of the LJ potential are listed in table~\ref{tab:parameters}. 

\begin{table}
\caption{Parameters of the Lennard-Jones potential}
\label{tab:parameters}   
\begin{tabular}{ccc|ccc}
\hline\noalign{\smallskip}
 Type & $\epsilon$ (emV) & $\sigma$ (\AA) &  Type  &  $\epsilon$ (emV) & $\sigma$ (\AA) \\
\noalign{\smallskip}\hline\noalign{\smallskip}
C-S & 6.03 & 3.29 &  C-B & 3.28 & 3.45  \\
C-N & 4.06 & 3.40 & C-Mo & 1.24 & 3.82 \\
B-S & 7.55 & 3.29 & N-S & 9.33 & 3.25 \\
B-Mo & 1.55 & 3.83 & N-Mo & 1.92 & 3.78 \\
\noalign{\smallskip}\hline
\end{tabular}
\end{table}

\section{Results and discussion}

\subsection{Temperature influence}

\begin{figure}[tb]
  \begin{center}
    \scalebox{1}[1]{\includegraphics[width=16cm]{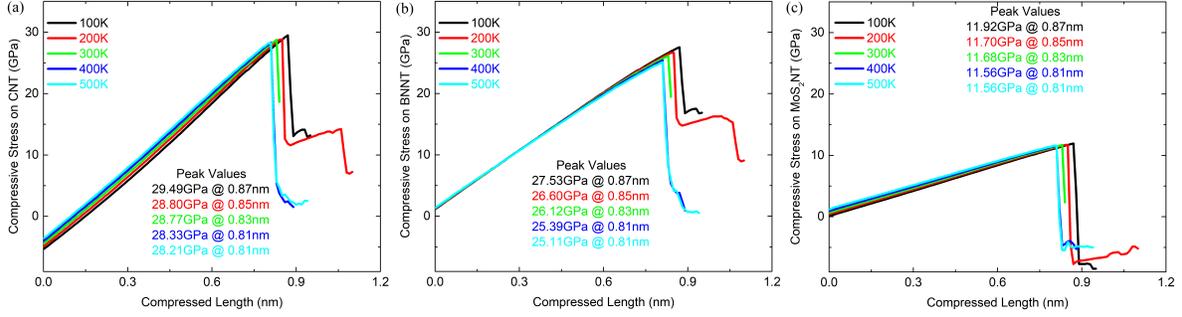}}
  \end{center}
  \caption{The histories of the axial stress of three nanotubes under compression test at different temperature.}
  \label{fig_temperature}
\end{figure}

In this composite heterostructure, the three nanotubes have different coefficients of thermal expansion. As both the ends of the composite heterostructure are fixed under the compression test, internal axial stress appears when the environment temperature is different from the synthesis temperature. We first discuss the temperature influence on the buckling behavior of the 1D vdWH. In this part, the CNT(20, 20)@BNNT(25, 25)@MoS$_2$NT(26, 26) 1D vdWH with an effective length of 24~nm is choosen as a sample to be uniaxially compressed. The environment temperature is set as 100K, 200K, 300K, 400K and 500K, respectively. 

From Fig.~\ref{fig_temperature}, one can find that all three nanotubes have non-zero initial axial stress. For example, in Fig.~\ref{fig_temperature}~(a), the initial values of stress of BNNT are 1.15GPa(100K), 1.21GPa(200K), 1.25GPa(300K), 1.30GPa(
K) and 1.36GPa(500K), respectively. It should be noted that a positive value of the compressive stress means that the nanotube is under compression rather than tension. After fully relexed, due to the mismatch of the thermal expansion coefficients among these three nanotubes, the CNT is preferred to be tensioned, while BNNT and MoS$_2$NT tend to be compressed. 

During compression, for each of these three nanotubes, the compressive stress increases monotonically to the peak value, and then rapidly drops. With the environment temperature increasing, the peak value of compressive stress decreases. The reason for this phenomenon is that the higher the environment temperature, the more strenuous thermal vibration of atoms in 1D vdWH, which is more likely to cause the buckling of the shells. For CNT, the axial stress varies from a negative value to a positive value when the compressed length reaches 0.13nm (engineering strain 0.5\%) at 300K, indicating a change in stress state of CNT from tension to compression. 


At the same temperature, as shown in Fig.~\ref{fig_temperature}, CNT has the largest peak value of compressive stress, while the smallest peak value of compressive stress is occured in MoS$_2$NT. The three critical compressed lengths corresponding to the three peak values of compressive stress are very close, $e.g$., the difference of compressed length is less than 0.001nm (engineering strain 0.04\%). It can be concluded that one nanotube buckles first, it will quickly trigger the other nanotubes to be buckled. The van der Waals interlayer interactions among the three shells play an important role in this deformation behavior. As one nanotube buckles, the radial deformation of the nanotube forces the other nanotubes to have the similar radial deformation, then rapidly buckle, and following this composite 1D vdWH damages as the bonds are seriously broken.


\subsection{Slenderness effects}

Slenderness, the ratio of effective length to radius of gyration of a column, is critical for evaluating the buckling behavior of colunms. To investigate the effect of slenderness on the buckling behavior of 1D vdWH, a varing slenderness ratio $\alpha$ is choosen as 3.6, 4.8, 6.0 and 7.2, respectively. Correspondingly, all the five 1D vdWHs have the same diameter 5~nm, i.e, the diameter of CNT(20, 20)@BNNT(25, 25)@MoS$_2$NT(26, 26) structure, and the effective lengths are 18nm, 24nm, 30nm and 36nm, respectively. 

\begin{figure}[tb]
  \begin{center}
    \scalebox{1}[1]{\includegraphics[width=16cm]{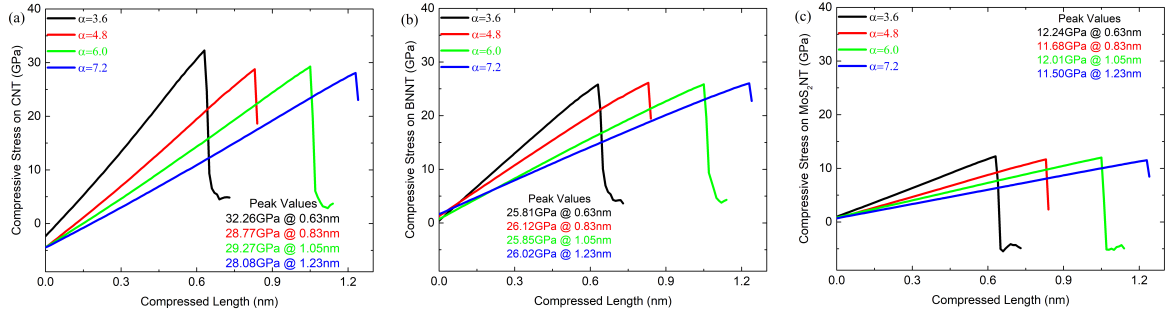}}
  \end{center}
  \caption{The histories of stress of three nanotubes in 1D vdWH under compression test with diffeent slenderness ratio.}
  \label{fig_slenderness}
\end{figure}

Fig.~\ref{fig_slenderness}~(a) illustrates the history of compressive stress of CNT during axial compression. The peak values of compressive stress in CNT have slight differences among the four cases, and the maximal value 32.25GPa is occured when $\alpha$ is 3.6. The critical strain corresponding to the peak values of compressive stress 3.27\% ($\alpha$=3.6), 3.30\%($\alpha$=4.8), 3.37\% ($\alpha$=6.0) and 3.31\% ($\alpha$=7.2), respectively. The values of critical strain of different $\alpha$ have slight difference. The similar results of BNNT and MoS$_2$NT can be found in Fig.~\ref{fig_slenderness}~(b) and Fig.~\ref{fig_slenderness}~(c). Hence, the slenderness ratio (4.8$\leq$ $\alpha$ $\leq$7.2) has a minor effect on the stability and strength of the 1D vdWH under axial compression. 

\section{conclusion}
In summary, we have investigated the buckling behavior of 1D vdWH under axial compression by using classical molecular dynamics. We find that as the temperature increases, the 1D vdWH gradully loses its stability and the peak value of compressive stress decreases. Slenderness ratio have a slight influence on the strength and stability of 1D vdWH under axial compression.

\textbf{Acknowledgment} The numerical calculations in this paper have been done on the National Supercomputing Center in Zhengzhou, and this work is supported by the Initial Founing of Scientific Rsearch of Zhengzhou University.

\bibliographystyle{aipnum4-1}
%

\end{document}